\documentclass[11pt,superscriptaddress,aps,prd,preprint]{revtex4}
\usepackage{amsmath}

\makeatletter
\usepackage[T1]{fontenc}
\usepackage{amsmath}
\usepackage{amssymb}
\usepackage{graphicx}

\usepackage{slashed}

\newcommand{\bea}{\begin{eqnarray}}
\newcommand{\eea}{\end{eqnarray}}

\newcommand{\e}{\epsilon}

\def\sumint{\iint\!\!\!\!\!\!\!\!\hbox{$\sum$}}

\begin{document}

\title{Quantized gravitoelectromagnetism theory at finite temperature}

\author{A. F. Santos}\email[]{alesandroferreira@fisica.ufmt.br}
\affiliation{Instituto de F\'{\i}sica, Universidade Federal de Mato Grosso,\\
78060-900, Cuiab\'{a}, Mato Grosso, Brazil}
\affiliation{Department of Physics and Astronomy, University of Victoria,\\
3800 Finnerty Road Victoria, BC, Canada}

\author{Faqir C. Khanna\footnote{Professor Emeritus - Physics Department, Theoretical Physics Institute, University of Alberta\\
Edmonton, Alberta, Canada}}\email[]{khannaf@uvic.ca}
\affiliation{Department of Physics and Astronomy, University of Victoria,\\
3800 Finnerty Road Victoria, BC, Canada}

\begin{abstract}

The Gravitoelectromagnetism (GEM) theory is considered in a lagrangian formulation using the Weyl tensor components. A perturbative approach to calculate processes at zero temperature has been used. Here the GEM at finite temperature is analyzed using Thermo Field Dynamics, real time finite temperature quantum field theory. Transition amplitudes involving gravitons, fermions and photons are calculated for various processes. These amplitudes are likely of interest in astrophysics.

\end{abstract}

\maketitle

\section{Introduction}

The unified theory for particle physics includes strong, weak and electromagnetic interactions. Experiments upto $\sim 2\, TeV$ are consistent with such as theory. The similarity between Newton's law and Coulomb's law lead Maxwell \cite{Maxwell} to formulate a theory of gravitation.  In similar vein Heaviside  \cite{Heaviside1} \cite{Heaviside2} developed equations for gravity. Based on these ideas the theory of Gravitoelectromagnetism (GEM) \cite{Thirring, Matte, Campbell1, Campbell2, Campbell3, Campbell4, Braginsky} was developed. GEM with group theoretical methods has been studied \cite{Jair1, Jair2}. The effects of the gravitomagnetic field on test particles in orbital motion in the slow-rotation regime have been analysed \cite{Lense}. There are numerous experiments to detect the gravitomagnetic contribution even though it is small \cite{Wheeler, Nordtvedt, Soffel, Everitt}. GEM has been analyzed with three different viewpoints: (i) the first is based on the similarity between the linearized Einstein and Maxwell equations \cite{Mashhon}; (ii) the second is based on the tidal tensors of the two theories \cite{Filipe} and (iii) the third is based on the Weyl tensor that is split into two parts \cite{Maartens}: electric and magnetic components. In this paper we will use the Weyl tensor approach with the Weyl tensor components ($C_{ijkl}$) being: ${\cal E}_{ij}=-C_{0i0j}$ (gravitoelectric field) and ${\cal B}_{ij}=\frac{1}{2}\epsilon_{ikl}C^{kl}_{0j}$ (gravitomagnetic field). The field equations for the components of the Weyl tensor have a structure similar to Maxwell equations.

Considering a symmetric gravitoelectromagnetic tensor potential $A_{\mu\nu}$, as the fundamental field which describes the gravitational interaction, a lagrangian formulation for GEM is constructed \cite{Khanna}. Using this formulation the interaction of gravitons with fermions and photons has been studied. Here the lagrangian, a real time quantum field theory, for GEM is considered at finite temperature using the Thermo Field Dynamics (TFD) formalism \cite{Umezawa1, Umezawa2, Umezawa22, Khanna1, Khanna2}. Its basic elements are the doubling of the original Fock space and using the Bogoliubov transformation. This doubling consists of Fock space composed of the original and a fictitious space (tilde space). The original and tilde space are related by a mapping, tilde conjugation rules. The Bogoliubov transformation is a rotation involving these two spaces. As a consequence the propagator is written in two parts: $T = 0$ and $T\neq 0$ components.

This paper is organized as follows. In section II, the lagrangian formulation of GEM is given. In section III, some characteristics of TFD are discussed. In section IV, the lagrangian formulation of GEM with TFD is analyzed and propagators for photon, fermions and graviton at finite temperature are presented. In section V, transition amplitudes of various processes at $T\neq 0$ are calculated. In section VI, some concluding remarks are made.

\section{Lagrangian formulation of GEM}

Here a brief introduction of the lagrangian formulation of GEM \cite{Khanna} is presented. This formulation is based on Maxwell-like equations
\bea
\partial^i{\cal E}^{ij}&=&-4\pi G\rho^j,\label{01}\\
\partial^i{\cal B}^{ij}&=&0,\label{02}\\
\epsilon^{\langle ikl}\partial^k{\cal B}^{lj\rangle}+\frac{1}{c}\frac{\partial{\cal E}^{ij}}{\partial t}&=&-\frac{4\pi G}{c}J^{ij},\label{03}\\
\epsilon^{\langle ikl}\partial^k{\cal E}^{lj\rangle}+\frac{1}{c}\frac{\partial{\cal B}^{ij}}{\partial t}&=&0,\label{04}
\eea
where $G$ is the gravitational constant, $\epsilon^{ikl}$ is the Levi-Civita symbol, $\rho^j$ is the vector mass density, $J^{ij}$ is the mass current density and $c$ is the speed of light. The gravitoelectric field ${\cal E}^{ij}$, the gravitomagnetic field ${\cal B}^{ij}$ and the mass current density $J^{ij}$ are symmetric traceless tensors of rank two. The symbol $\langle\cdots\rangle$ denotes symmetrization of the first and last indices i.e. $i$ and $j$.

The fields ${\cal E}^{ij}$ and ${\cal B}^{ij}$ are expressed in terms of a symmetric rank-2 tensor field, $\tilde{\cal A}$, with components ${\cal A}^{ij}$, such that
\bea
{\cal B}=\textrm{curl}\,\tilde{\cal A},
\eea
with ${\cal B}^{ij}=\epsilon^{\langle ikl}\partial^k{\cal A}^{lj\rangle}$. To satisfy eq. (\ref{02}), $\textrm{div}\, \textrm{curl}\,\tilde{\cal A}=\frac{1}{2}\mathrm{curl}\,\mathrm{div}\,\tilde{\cal A}$ has been used and $\tilde{\cal A}$ is such that $\mathrm{div}\,\tilde{\cal A}=0$. With $\mathrm{curl}\,{\cal E}=\epsilon^{\langle ikl}\partial^k{\cal E}^{lj\rangle}$ it is possible to rewrite eq. (\ref{04}) as
\bea
\mathrm{curl}\left({\cal E}+\frac{1}{c}\frac{\partial\tilde{\cal A}}{\partial t}\right)=0.
\eea
Then the gravitoelectric field is
\bea
{\cal E}+\frac{1}{c}\frac{\partial\tilde{\cal A}}{\partial t}=-\mathrm{grad}\,\varphi.
\eea
Here $\varphi$ is the GEM counterpart of the electromagnetic (EM) scalar potential $\phi$. Thus the GEM fields ${\cal E}$ and ${\cal B}$ are defined as
\bea
{\cal E}&=&-\mathrm{grad}\,\varphi-\frac{1}{c}\frac{\partial \tilde{\cal A}}{\partial t},\\
{\cal B}&=&\mathrm{curl}\,\tilde{\cal A}.
\eea

The GEM fields are elements of a rank-3 tensor, gravitoelectromagnetic tensor ${\cal F}^{\mu\nu\alpha}$, 
\bea
{\cal F}^{\mu\nu\alpha}=\partial^\mu{\cal A}^{\nu\alpha}-\partial^\nu{\cal A}^{\mu\alpha},
\eea
where $\mu, \nu,\alpha=0, 1, 2, 3$. The non-zero components of ${\cal F}^{\mu\nu\alpha}$ are
\bea
{\cal F}^{0ij}&=&{\cal E}^{ij},\\
{\cal F}^{ijk}&=&\epsilon^{ijl}{\cal B}^{lk}.
\eea

The dual GEM tensor is
\bea
{\cal G}^{\mu\nu\alpha}=\frac{1}{2}\epsilon^{\mu\nu\gamma\sigma}\eta^{\alpha\beta}{\cal F}_{\gamma\sigma\beta},
\eea
where ${\cal F}_{\gamma\sigma\beta}=\eta_{\gamma\mu}\eta_{\sigma\nu}\eta_{\beta\alpha}{\cal F}^{\mu\nu\alpha}$ and $\eta_{\mu\nu}=(+,-,-,-)$. 

The Maxwell-like equations are written as
\bea
\partial_\mu{\cal F}^{\mu\nu\alpha}&=&\frac{4\pi G}{c}{\cal J}^{\nu\alpha},\\
\partial_\mu{\cal G}^{\mu\langle\nu\alpha\rangle}&=&0,
\eea
where ${\cal J}^{\nu\alpha}$ is a rank-2 tensor that depends on the mass density $\rho^i$ and the current density $J^{ij}$. With these ingredients the GEM lagrangian density is written as
\bea
{\cal L}_G=-\frac{1}{16\pi}{\cal F}_{\mu\nu\alpha}{\cal F}^{\mu\nu\alpha}-\frac{G}{c}\,{\cal J}^{\nu\alpha}{\cal A}_{\nu\alpha}.\label{L_G}
\eea
The quantisation of GEM with the symmetric tensor ${\cal A}_{\nu\alpha}$ leads to spin-2 gravitons \cite{Khanna}, in analogy to the electromagnetism field where the vector potential $A^\mu$  yields spin-1 photons.

The lagrangian density of GEM \cite{Khanna} including interactions of gravitons with photons and fermions is given as 
\bea
{\cal L}={\cal L}_G+{\cal L}_F+{\cal L}_A+{\cal L}_{FA}+{\cal L}_{GF}+{\cal L}_{GA}+{\cal L}_{GFA},\label{Full}
\eea
where ${\cal L}_G$ is given in eq. (\ref{L_G}). The fermion field is described by 
\bea
{\cal L}_F=-\frac{i\hbar c}{2}\left(\bar{\psi}\gamma^\mu\partial_\mu\psi-\partial_\mu\bar{\psi}\gamma^\mu\psi\right)+mc^2\bar{\psi}\psi,
\eea
with $\psi$ being the fermion field and $\bar{\psi}=\psi^\dagger \gamma_0$. For the EM field the lagrangian is
\bea
{\cal L}_A=-\frac{1}{16\pi} F_{\mu\nu}F^{\mu\nu},
\eea
where $F_{\mu\nu}=\partial_\mu A_\nu-\partial_\nu A_\mu$ and $A_\mu$ is the vector potential. The interaction between the fermion field and $A_\mu$ is given by
\bea
{\cal L}_{FA}=e\bar{\psi}\gamma^\mu\psi A_\mu,
\eea
where $e$ is the coupling constant. The interaction between ${\cal A}_{\mu\nu}$ and the fermion field is described by
\bea
{\cal L}_{GF}=-\frac{i\hbar c\kappa}{4}{\cal A}_{\mu\nu}\left(\bar{\psi}\gamma^\mu\partial^\nu\psi-\partial^\mu\bar{\psi}\gamma^\nu\psi\right),
\eea
with $\kappa=\frac{\sqrt{8\pi G}}{c^2}$ being the coupling constant. Now the interaction between photon and graviton is given by
\bea
{\cal L}_{GA}=\frac{\kappa}{4\pi}{\cal A}_{\mu\nu}\left(F^\mu_\alpha F^{\nu\alpha}-\frac{1}{4}\eta^{\mu\nu}F^{\alpha\rho}F_{\alpha\rho}\right),
\eea
and the interaction between the photon, graviton and fermion is 
\bea
{\cal L}_{GFA}=\frac{1}{2}e\kappa\bar{\psi}\gamma^\mu\psi A^\nu {\cal A}_{\mu\nu}.
\eea

Our aim here is to describe this theory at finite temperature using the TFD formalism. Details such as gauge invariance and equations of motion are given earlier \cite{Khanna}.

\section{Thermo Field Dynamics - TFD}

TFD is a formalism where the thermal average of an observable is given by the vacuum expectation value in an extended Fock space. This is obtained when a thermal ground state $|0(\beta)\rangle$ is constructed, where $\beta=\frac{1}{k_B T}$,  $k_B$ is the Boltzmann constant and T is the temperature. This formalism is constructed with basic two ingredients: (a) a doubling of the Fock space, ${\cal S}$, of the original field system, giving rise to ${\cal S}_T={\cal S}\otimes \tilde{\cal S}$, applicable to systems in  a thermal equilibrium state. This doubling is defined by the tilde conjugation rules, associating each operator say $a$, in ${\cal S}$ to two operators in ${\cal S}_T$, say
\bea
A=a\otimes 1,\quad\quad\quad\quad \tilde{A}=1\otimes a,
\eea
such that the physical quantities are described by the nontilde operators. (b) A Bogoliubov transformation that introduces a rotation in the tilde and nontilde variables. Then, the thermal quantities are introduced by a Bogoliubov transformation. 

\subsection{For bosons} 

Bogoliubov transformations for bosons are given as
\bea
a(k)&=&c_B(\omega)a(k,\beta)+d_B(\omega)\tilde a^\dagger(k,\beta),\nonumber\\
a^\dagger(k)&=&c_B(\omega)a^\dagger(k,\beta)+d_B(\omega)\tilde a(k,\beta),\nonumber\\
\tilde a(k)&=&c_B(\omega)\tilde a(k,\beta)+d_B(\omega) a^\dagger(k,\beta),\nonumber\\
\tilde a^\dagger(k)&=&c_B(\omega)\tilde a^\dagger(k,\beta)+d_B(\omega) a(k,\beta),\label{BTp}
\eea
where $(a^\dagger, \tilde{a}^\dagger)$ are creation operators and $(a, \tilde{a})$ are destruction operators, with
\bea
c_B^2(\omega)=1+f_B(\omega),\quad\quad d_B^2(\omega)=f_B(\omega), \quad\quad f_B(\omega)=\frac{1}{e^{\beta\omega}-1},\label{phdef}
\eea
where $\omega=\omega(k)$. 

Algebraic rules for thermal operators are
\bea
\left[a(k, \beta), a^\dagger(p, \beta)\right]=\delta^3(k-p),\quad\quad\quad \left[\tilde{a}(k, \beta), \tilde{a}^\dagger(p, \beta)\right]=\delta^3(k-p),\label{ComB}
\eea
and other commutation relations are null.

\subsection{For fermions}

Bogoliubov transformations for fermions are given as
\bea
a(k)&=&c_F(\omega)a(k,\beta)+d_F(\omega)\tilde a^\dagger(k,\beta),\nonumber\\
a^\dagger(k)&=&c_F(\omega)a^\dagger(k,\beta)+d_F(\omega)\tilde a(k,\beta),\nonumber\\
\tilde a(k)&=&c_F(\omega)\tilde a(k,\beta)-d_F(\omega) a^\dagger(k,\beta),\nonumber\\
\tilde a^\dagger(k)&=&c_F(\omega)\tilde a^\dagger(k,\beta)-d_F(\omega) a(k,\beta),\label{BTf}
\eea
with
\bea
c_F^2(\omega)=1-f_F(\omega),\quad\quad d_F^2(\omega)=f_F(\omega), \quad\quad f_F(\omega)=\frac{1}{e^{\beta\omega}+1}.\label{ferdef}
\eea

Algebraic rules for thermal operators are
\bea
\left\{a(k, \beta), a^\dagger(p, \beta)\right\}=\delta^3(k-p),\quad\quad\quad \left\{\tilde{a}(k, \beta), \tilde{a}^\dagger(p, \beta)\right\}=\delta^3(k-p),\label{ComF}
\eea
and other commutation relations are null.

In the next section the TFD formalism is used to write the GEM lagrangian at finite temperature.

\section{Quantized GEM at finite temperature}

Here the quantized GEM theory at finite temperature is considered. The doubled lagrangian $\hat{\cal L}$ is written as
\bea
\hat{\cal L}={\cal L}-\tilde{\cal L},
\eea
where ${\cal L}$ is the lagrangian of the physical system that includes interactions of gravitons with fermions and photons as given in eq. (\ref{Full}). The lagrangian $\tilde{\cal L}$ describes the tilde ($\thicksim$) system and is given by
\bea
\tilde{\cal L}=\tilde{\cal L}_G+\tilde{\cal L}_F+\tilde{\cal L}_A+\tilde{\cal L}_{FA}+\tilde{\cal L}_{GF}+\tilde{\cal L}_{GA}+\tilde{\cal L}_{GFA}, 
\eea
where
\bea
\tilde{\cal L}_G&=&-\frac{1}{16\pi}\tilde{\cal F}_{\mu\nu\alpha}\tilde{\cal F}^{\mu\nu\alpha},\\
\tilde{\cal L}_F&=&-\frac{i\hbar c}{2}\left(\bar{\tilde{\psi}}\gamma^\mu\partial_\mu\tilde{\psi}-\partial_\mu\bar{\tilde{\psi}}\gamma^\mu\tilde{\psi}\right)+mc^2\bar{\tilde{\psi}}\tilde{\psi},\\
\tilde{\cal L}_A&=&-\frac{1}{16\pi} \tilde F_{\mu\nu}\tilde F^{\mu\nu},\\
\tilde{\cal L}_{FA}&=&e\bar{\tilde \psi}\gamma^\mu\tilde \psi \tilde A_\mu,\\
\tilde{\cal L}_{GF}&=&-\frac{i\hbar c\kappa}{4}\tilde {\cal A}_{\mu\nu}\left(\bar{\tilde\psi}\gamma^\mu\partial^\nu\tilde \psi-\partial^\mu\bar{\tilde\psi}\gamma^\nu\tilde\psi\right),\\
\tilde{\cal L}_{GA}&=&\frac{\kappa}{4\pi}\tilde{\cal A}_{\mu\nu}\left(\tilde F^\mu_\alpha \tilde F^{\nu\alpha}-\frac{1}{4}\eta^{\mu\nu}\tilde F^{\alpha\rho}\tilde F_{\alpha\rho}\right),\\
\tilde{\cal L}_{GFA}&=&\frac{1}{2}e\kappa\bar{\tilde \psi}\gamma^\mu\tilde \psi \tilde A^\nu \tilde {\cal A}_{\mu\nu}.
\eea

Using this formalism the photon, electron and graviton propagator are obtained. The propagator is written in two parts: one describes the flat space-time contribution and the other displays the thermal and/or the topological effect.

\subsection{The Photon Propagator}

The photon propagator is \cite{Umezawa2}
\bea
i\Delta_{\mu\nu}(x-y)&=&\langle 0(\beta)|T\,(A_{\mu}(x)A_{\nu}(y))|0(\beta)\rangle\\
&=&\theta(t_x-t_y)\langle 0(\beta)|A_{\mu}(x)A_{\nu}(y)|0(\beta)\rangle + \theta(t_y-t_x)\langle 0(\beta)|A_{\nu}(y)A_{\mu}(x)|0(\beta)\rangle\nonumber,
\eea
where the vector $A_\mu(x)$ is given by
\bea
A_{\mu}(x)=\int\frac{d^3k}{\sqrt{2\omega_k(2\pi)^3}}\sum_\lambda\epsilon_{\mu}(k,\lambda)\left(a_{k,\lambda}e^{-ik_\rho x^\rho}+a_{k,\lambda}^\dagger e^{ik_\rho x^\rho}\right),\label{A1}
\eea
with $\epsilon_{\mu}(k,\lambda)$ being the polarization vector. This propagator is represented in FIG. 1.
\begin{figure}[h]
\includegraphics[scale=0.4]{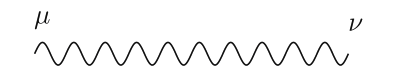}
\caption{Photon Propagator}
\end{figure}

The two point function in TFD is a thermal doublet, and has $2\times2$ matrix structure
\bea
\left( \begin{array}{c} A^1_{\mu} \\ A^2_{\mu} \end{array} \right)=\left( \begin{array}{c} A_{\mu} \\ {\tilde A_{\mu}^\dagger} \end{array} \right).\label{doublet1}
\eea
Then the photon propagator is 
\bea
i\Delta^{ab}_{\mu\nu}(x-y)=\theta(t_x-t_y)\langle 0(\beta)|A^a_{\mu}(x)A^b_{\nu}(y)|0(\beta)\rangle + \theta(t_y-t_x)\langle 0(\beta)|A^b_{\nu}(y)A^a_{\mu}(x)|0(\beta)\rangle,\label{Delta}
\eea
where $a,b=1,2$ and $\mu,\,\nu$ are tensor indices. Using the Cauchy theorem
\bea
\int dk^0\frac{e^{-ik_0(x_0-y_0)}}{k_0-(\omega-i\xi)}&=&-2\pi ie^{-i\omega_k(x_0-y_0)}\theta(x_0-y_0),\nonumber\\
\int dk^0\frac{e^{-ik_0(x_0-y_0)}}{k_0-(-\omega+i\xi)}&=&2\pi ie^{i\omega_k(x_0-y_0)}\theta(y_0-x_0),\label{Cauchy}
\eea
then for $a=b=1$ eq. (\ref{Delta}) becomes
\bea
\Delta^{11}_{\mu\nu}(x-y)=-\int\frac{d^4k}{(2\pi)^4}e^{-ik_\rho(x^\rho -y^\rho)}\sum_{\lambda}\epsilon_{\mu}(k,\lambda)\epsilon_{\nu}(k,\lambda)\left[\frac{c_B^2(\omega)}{k_0^2-(\omega-i\xi)^2}-\frac{d_B^2(\omega)}{k_0^2-(\omega+i\xi)^2}\right].\label{11}
\eea
Here $k_\rho(x^\rho -y^\rho)\equiv ik_0(t_x-t_y)-\vec{k}\cdot(\vec{x}-\vec{y})$. Calculating other components $\Delta^{12}_{\mu\nu}(x-y)$, $\Delta^{21}_{\mu\nu}(x-y)$ and $\Delta^{22}_{\mu\nu}(x-y)$, the propagator is 
\bea
\Delta^{ab}_{\mu\nu}(x-y)=\frac{i\hbar}{(2\pi)^4}\int d^4k\,e^{ik\cdot(x-y)-ik_0(t_x-t_y)}\Delta^{ab}_{\mu\nu}(k)
\eea
and
\bea
\Delta^{ab}_{\mu\nu}(k)=\left( \begin{array}{cc} \frac{c_B^2(\omega)}{k_0^2-(\omega-i\xi)^2}-\frac{d_B^2(\omega)}{k_0^2-(\omega+i\xi)^2} \hspace{0,2cm} & \hspace{0,2cm} \frac{c_B(\omega)d_B(\omega)}{k_0^2-(\omega-i\xi)^2}-\frac{c_B(\omega)d_B(\omega)}{k_0^2-(\omega+i\xi)^2} \\ 
\frac{c_B(\omega)d_B(\omega)}{k_0^2-(\omega-i\xi)^2}-\frac{c_B(\omega)d_B(\omega)}{k_0^2-(\omega+i\xi)^2} \hspace{0,2cm} & \hspace{0,2cm} \frac{d_B^2(\omega)}{k_0^2-(\omega-i\xi)^2}-\frac{c_B^2(\omega)}{k_0^2-(\omega+i\xi)^2} \end{array} \right)\eta_{\mu\nu},\label{PH}
\eea
with $\omega\equiv\omega(k)$ and $\sum_{\lambda}\epsilon_{\mu}(k,\lambda)\epsilon_{\nu}(k,\lambda)=\eta_{\mu\nu}$. In a simplified form eq. (\ref{PH}) is given as
\bea
\Delta^{ab}_{\mu\nu}(k)=U_B(\omega)\tau\left[k_0^2-(\omega-i\delta\tau)^2\right]^{-1}U_B(\omega)\eta_{\mu\nu}, 
\eea
where
\bea
U_B(\omega)=\left( \begin{array}{cc}c_B(\omega) & d_B(\omega) \\ 
d_B(\omega) & c_B(\omega)\end{array} \right), \quad\quad\quad \tau=\left( \begin{array}{cc}1 & 0 \\ 
0 & -1\end{array} \right).\label{UB}
\eea

The propagator is separated as
\bea
\Delta_{\mu\nu}(k)=\Delta_{\mu\nu}^{(0)}(k)+\Delta_{\mu\nu}^{(\beta)}(k),\label{photon1}
\eea
where $\Delta_{\mu\nu}^{(0)}(k)$ and $\Delta_{\mu\nu}^{(\beta)}(k)$ are zero and finite temperature parts respectively. Explicitly 
\bea
\Delta_{\mu\nu}^{(0)}(k)&=&\frac{\eta_{\mu\nu}}{k^2}\tau,\nonumber\\
\Delta_{\mu\nu}^{(\beta)}(k)&=&-\frac{2\pi i\delta(k^2)}{e^{\beta k_0}-1}\left( \begin{array}{cc}1&e^{\beta k_0/2}\\e^{\beta k_0/2}&1\end{array} \right)\eta_{\mu\nu}.
\eea

\subsection{The Electron Propagator}

The electron propagator is given as
\bea
S^{ab}_{\mu\nu}(x-y)=\hbar\int\frac{d^4k}{(2\pi)^4}e^{-ik_\rho(x^\rho-y^\rho)}S^{ab}_{\mu\nu}(k),
\eea
and it is represented in FIG. 2.
\begin{figure}[h]
\includegraphics[scale=0.4]{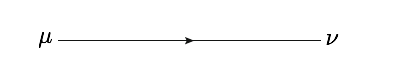}
\caption{Fermion Propagator}
\end{figure}

This equation is written as
\bea
S^{ab}_{\mu\nu}(k)=\frac{1}{\hbar}\int d^4z\,e^{ik_\rho(x^\rho-y^\rho)}S^{ab}_{\mu\nu}(x-y),\label{electron}
\eea
where $z=x-y$ and
\bea
S^{ab}_{\mu\nu}(x-y)=\theta(t_x-t_y)\langle 0(\beta)|\psi^a_{\mu}(x)\bar{\psi}^b_{\nu}(y)|0(\beta)\rangle - \theta(t_y-t_x)\langle 0(\beta)|\bar{\psi}^b_{\nu}(y)\psi^a_{\mu}(x)|0(\beta)\rangle,\label{electron2}
\eea
with $\bar{\psi}(y)=\psi^\dagger(y)\gamma^0$ and $\mu, \nu$ being the spinors indices. Using the free Dirac field equation,
\bea
\psi(x)=\sum_r\int d^3k\, \left[u^r(k)a^r(k)e^{i(k\cdot x-\xi_k t)}+v^r(k)b^{r\dagger}(k)e^{-i(k\cdot x-\xi_k t)}\right],
\eea
where $a^r(k)$ and $b^{r}(k)$ are annihilation operators for electrons and positrons, respectively. $u^r(k)$ and $v^r(k)$ are Dirac spinors then eq. (\ref{electron2}) becomes
\bea
S^{ab}_{\mu\nu}(k)&=&\left(\frac{\gamma^0\xi-\vec{\gamma}\cdot \vec{k}+m}{2\xi}\right)_{\mu\nu}\left[U_F(\xi)(k_0-\xi +i\delta\tau)^{-1}U_F^\dagger(\xi)\right]^{ab}\nonumber\\&+&\left(\frac{\gamma^0\xi+\vec{\gamma}\cdot \vec{k}-m}{2\xi}\right)_{\mu\nu}\left[U_F(-\xi)(k_0+\xi +i\delta\tau)^{-1}U_F^\dagger(-\xi)\right]^{ab}, 
\eea
where Bogoliubov transformations, eq. (\ref{BTf}), are used. Here $\xi\equiv\xi(k)$,
\bea
U_F(\xi)=\left( \begin{array}{cc}c_F(\xi) & d_F(\xi) \\ 
-d_F(\xi) & c_F(\xi)\end{array} \right),
\eea
and $c_F(\xi)$ and $d_F(\xi)$ are given by  eq. (\ref{ferdef}).

This propagator is separated into two parts as
\bea
S_{\mu\nu}(k)=S_{\mu\nu}^{(0)}(k)+S_{\mu\nu}^{(\beta)}(k),\label{fermion}
\eea
where
\bea
S_{\mu\nu}^{(0)}(k)&=&\frac{\slashed k +m}{k^2-m^2},\nonumber\\
S_{\mu\nu}^{(\beta)}(k)&=&\frac{2\pi i}{e^{\beta k_0}+1}\Biggl[\frac{\gamma^0\epsilon-\vec{\gamma}\cdot\vec{k}+m}{2\epsilon}\left( \begin{array}{cc}1&e^{\beta k_0/2}\\e^{\beta k_0/2}&-1\end{array} \right)\delta(k_0-\epsilon)\nonumber\\
&+&\frac{\gamma^0\epsilon+\vec{\gamma}\cdot\vec{k}-m}{2\epsilon}\left( \begin{array}{cc}-1&e^{\beta k_0/2}\\e^{\beta k_0/2}&1\end{array} \right)\delta(k_0+\epsilon)\Biggl].
\eea
Here $S_{\mu\nu}^{(0)}(k)$ and $S_{\mu\nu}^{(\beta)}(k)$ are zero and finite temperature parts respectively.

\subsection{The Graviton Propagator}

The graviton propagator, represented in FIG. 3, 
\begin{figure}[h]
\includegraphics[scale=0.4]{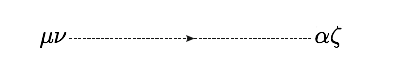}
\caption{Graviton Propagator}
\end{figure}
is written as
\bea
iD^{ab}_{\mu\nu\alpha\zeta}(x-y)=\theta(t_x-t_y)\langle 0(\beta)|A^a_{\mu\nu}(x)A^b_{\alpha\zeta}(y)|0(\beta)\rangle + \theta(t_y-t_x)\langle 0(\beta)|A^b_{\alpha\zeta}(y)A^a_{\mu\nu}(x)|0(\beta)\rangle,
\eea
where $a, b=1, 2$ and $\mu,\,\nu,\,\alpha,\,\zeta$ are tensor indices. The tensor $A_{\mu\nu}(x)$ is given by
\bea
A_{\mu\nu}(x)=\int\frac{d^3k}{\sqrt{2\omega_k(2\pi)^3}}\sum_\lambda\epsilon_{\mu\nu}(k,\lambda)\left(a_{k,\lambda}e^{-ik_\rho x^\rho}+a_{k,\lambda}^\dagger e^{ik_\rho x^\rho}\right),\label{A}
\eea
with $\epsilon_{\mu\nu}(k,\lambda)$ being the polarization tensor. Using the thermal doublet,
\bea
\left( \begin{array}{c} A^1_{\mu\nu} \\ A^2_{\mu\nu} \end{array} \right)=\left( \begin{array}{c} A_{\mu\nu} \\ {\tilde A_{\mu\nu}^\dagger} \end{array} \right),\label{doublet}
\eea
the component $a=b=1$ is given as
\bea
iD^{11}_{\mu\nu\alpha\zeta}(x-y)&=&\theta(t_x-t_y)\sumint dp\,dk\,\epsilon_{\mu\nu}(k,\lambda)\epsilon_{\alpha\zeta}(p,\lambda')\times\nonumber\\
&\times &\langle 0(\beta)| \left(a_{k,\lambda}e^{-ik_\rho x^\rho}+a_{k,\lambda}^\dagger e^{ik_\rho x^\rho}\right) \left(a_{p,\lambda}e^{-ip_\rho y^\rho}+a_{p,\lambda}^\dagger e^{ip_\rho y^\rho}\right)|0(\beta)\rangle\nonumber\\
&+&\theta(t_y-t_x)\sumint dp\,dk\,\epsilon_{\alpha\zeta}(p,\lambda')\epsilon_{\mu\nu}(k,\lambda)\times\nonumber\\
&\times &\langle 0(\beta)| \left(a_{p,\lambda}e^{-ip_\rho y^\rho}+a_{p,\lambda}^\dagger e^{ip_\rho y^\rho}\right)\left(a_{k,\lambda}e^{-ik_\rho x^\rho}+a_{k,\lambda}^\dagger e^{ik_\rho x^\rho}\right)|0(\beta)\rangle,
\eea
where $\sumint dp\,dk\equiv \int\frac{d^3k}{\sqrt{2\omega_k(2\pi)^3}}\int\frac{d^3p}{\sqrt{2\omega_p(2\pi)^3}}\sum_{\lambda,\lambda'}$ has been used. Using Bogoliubov transformations, eq. (\ref{BTp}), and commutation relations, eq. (\ref{ComB}), the propagator is
\bea
iD^{11}_{\mu\nu\alpha\zeta}(x-y)&=&-\theta(t_x-t_y)\sumint dp\,dk\,\epsilon_{\mu\nu}(k,\lambda)\epsilon_{\alpha\zeta}(p,\lambda')\times\nonumber\\
&\times &\left[c_B^2(\omega)\delta^3(k-p)e^{-ik_\rho x^\rho+ip_\rho y^\rho}+d_B^2(\omega)\delta^3(k-p)e^{ik_\rho x^\rho-ip_\rho y^\rho}\right]\nonumber\\
&-&\theta(t_y-t_x)\sumint dp\,dk\,\epsilon_{\alpha\zeta}(p,\lambda')\times\nonumber\\
&\times &\left[c_B^2(\omega)\delta^3(p-k)e^{-ip_\rho y^\rho+ik_\rho x^\rho}+d_B^2(\omega)\delta^3(p-k)e^{ip_\rho y^\rho-ik_\rho x^\rho}\right].
\eea
Applying the Cauchy theorem, eq. (\ref{Cauchy}), we get 
\bea
D^{11}_{\mu\nu\alpha\zeta}(x-y)=-\int\frac{d^4k}{(2\pi)^4}e^{-ik_\rho(x^\rho -y^\rho)}\left[\frac{c_B^2(\omega)}{k_0^2-(\omega-i\xi)^2}-\frac{d_B^2(\omega)}{k_0^2-(\omega+i\xi)^2}\right]\varepsilon_{\mu\nu\alpha\zeta},\label{1}
\eea
with
\bea
\sum_{\lambda}\epsilon_{\mu\nu}(k,\lambda)\epsilon_{\alpha\zeta}(k,\lambda)=\frac{1}{2}\left(\eta_{\mu\alpha}\eta_{\nu\zeta}+\eta_{\mu\zeta}\eta_{\nu\alpha}-\eta_{\mu\nu}\eta_{\alpha\zeta}\right)\equiv \varepsilon_{\mu\nu\alpha\zeta}.
\eea

The component with $a=1,\,b=2$ is written as
\bea
iD^{12}_{\mu\nu\alpha\zeta}(x-y)=\theta(t_x-t_y)\langle 0(\beta)|A_{\mu\nu}(x)\tilde A_{\alpha\zeta}^\dagger(y)|0(\beta)\rangle + \theta(t_y-t_x)\langle 0(\beta)|\tilde A_{\alpha\zeta}^\dagger(y)A_{\mu\nu}(x)|0(\beta)\rangle,
\eea
where the doublet notation is used. This component is written as
\bea
D^{12}_{\mu\nu\alpha\zeta}(x-y)=-\int\frac{d^4k}{(2\pi)^4}e^{-ik_\rho(x^\rho -y^\rho)}\left[\frac{c_B(\omega)d_B(\omega)}{k_0^2-(\omega-i\xi)^2}-\frac{c_B(\omega)d_B(\omega)}{k_0^2-(\omega+i\xi)^2}\right]\varepsilon_{\mu\nu\alpha\zeta}.\label{2}
\eea
In addition we have
\bea
D^{21}_{\mu\nu\alpha\zeta}(x-y)=D^{12}_{\mu\nu\alpha\zeta}(x-y). \label{21}
\eea

For the component $a=b=2$, we get 
\bea
iD^{22}_{\mu\nu\alpha\zeta}(x-y)&=&\theta(t_x-t_y)\langle 0(\beta)|\tilde A_{\mu\nu}^\dagger(x)\tilde A_{\alpha\zeta}^\dagger(y)|0(\beta)\rangle + \theta(t_y-t_x)\langle 0(\beta)|\tilde A_{\alpha\zeta}^\dagger(y)\tilde A_{\mu\nu}^\dagger(x)|0(\beta)\rangle,\nonumber\\
&=&-\int\frac{d^4k}{(2\pi)^4}e^{-ik_\rho(x^\rho -y^\rho)}\left[\frac{d_B^2(\omega)}{k_0^2-(\omega-i\xi)^2}-\frac{c_B^2(\omega)}{k_0^2-(\omega+i\xi)^2}\right]\varepsilon_{\mu\nu\alpha\zeta}.\label{3}
\eea

Combining eqs. (\ref{1}), (\ref{2}), (\ref{21}) and (\ref{3}) the graviton propagator is
\bea
D^{ab}_{\mu\nu\alpha\zeta}(x-y)=-\int\frac{d^4k}{(2\pi)^4}e^{-ik_\rho(x^\rho -y^\rho)} \,D^{ab}_{\mu\nu\alpha\zeta}(k),
\eea
where
\bea
D^{ab}_{\mu\nu\alpha\zeta}(k)=\left( \begin{array}{cc} \frac{c_B^2(\omega)}{k_0^2-(\omega-i\xi)^2}-\frac{d_B^2(\omega)}{k_0^2-(\omega+i\xi)^2} \hspace{0,2cm} & \hspace{0,2cm} \frac{c_B(\omega)d_B(\omega)}{k_0^2-(\omega-i\xi)^2}-\frac{c_B(\omega)d_B(\omega)}{k_0^2-(\omega+i\xi)^2} \\ 
\frac{c_B(\omega)d_B(\omega)}{k_0^2-(\omega-i\xi)^2}-\frac{c_B(\omega)d_B(\omega)}{k_0^2-(\omega+i\xi)^2} \hspace{0,2cm} & \hspace{0,2cm} \frac{d_B^2(\omega)}{k_0^2-(\omega-i\xi)^2}-\frac{c_B^2(\omega)}{k_0^2-(\omega+i\xi)^2} \end{array} \right)\,\varepsilon_{\mu\nu\alpha\zeta}.\label{4}
\eea

Then eq. (\ref{4}) is written in the following form:
\bea
D^{ab}_{\mu\nu\alpha\zeta}(k)=U_B(\omega)\tau\left[k_0^2-(\omega-i\delta\tau)^2\right]^{-1}U_B(\omega)\varepsilon_{\mu\nu\alpha\zeta},
\eea
where $U_B(\omega)$ and $\tau$ are given in eq. (\ref{UB}).

The graviton propagator is written as
\bea
D_{\mu\nu\alpha\zeta}(k)=D_{\mu\nu\alpha\zeta}^{(0)}(k)+D_{\mu\nu\alpha\zeta}^{(\beta)}(k),\label{graviton}
\eea
with
\bea
D_{\mu\nu\alpha\zeta}^{(0)}(k)&=& \frac{\eta_{\mu\alpha}\eta_{\nu\zeta}+\eta_{\mu\zeta}\eta_{\nu\alpha}-\eta_{\mu\nu}\eta_{\alpha\zeta}}{2k^2}\,\tau,\nonumber\\
D_{\mu\nu\alpha\zeta}^{(\beta)}(k)&=&-\frac{\pi i\delta(k^2)}{e^{\beta k_0}-1}\left( \begin{array}{cc}1&e^{\beta k_0/2}\\e^{\beta k_0/2}&1\end{array} \right)(\eta_{\mu\alpha}\eta_{\nu\zeta}+\eta_{\mu\zeta}\eta_{\nu\alpha}-\eta_{\mu\nu}\eta_{\alpha\zeta}),
\eea
where $D_{\mu\nu\alpha\zeta}^{(0)}(k)$ and $D_{\mu\nu\alpha\zeta}^{(\beta)}(k)$ are zero  and finite temperature parts respectively.

\section{Gravitation interacting with photons and fermions at finite temperature}

Transition amplitudes, ${\cal M}$, of various scattering processes involving gravitons, photons and fermions at finite temperature are calculated. The vertex factors are:

\begin{enumerate}
\item Graviton-Photon vertex factor
\begin{figure}[h]
\includegraphics[scale=0.4]{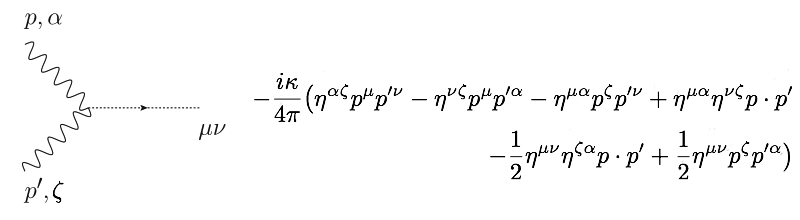}
\end{figure}

\newpage

\item Fermion-Photon vertex factor
\begin{figure}[h]
\includegraphics[scale=0.4]{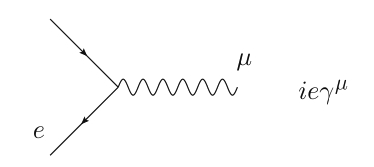}
\end{figure}

\item Graviton-Fermion vertex factor
\begin{figure}[h]
\includegraphics[scale=0.4]{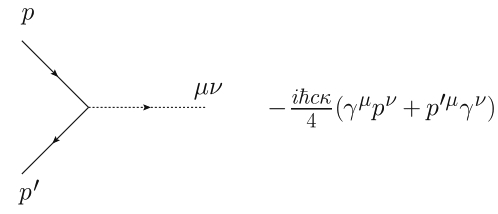}
\end{figure}

\item Graviton-Fermion-Photon vertex factor
\begin{figure}[h]
\includegraphics[scale=0.4]{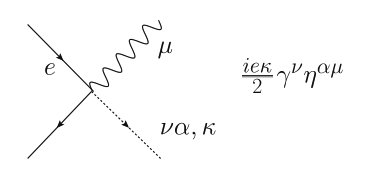}
\end{figure}
\end{enumerate}

\subsection{Gravitational M$\varnothing$ller scattering}

The diagrams are similar to the traditional M$\varnothing$ller scattering, where the photon is replaced by a graviton. The two diagrams that describe this scattering are given in FIG. 4.
\begin{figure}[htb]
\includegraphics[scale=0.35]{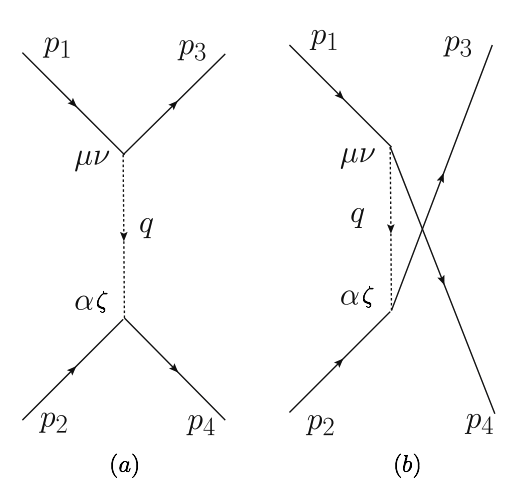}
\caption{Feynman diagram for M$\varnothing$ller scattering}
\end{figure}

The total transition amplitude is given by
\bea
{\cal M}= {\cal M}_1+{\cal M}_2,
\eea
where ${\cal M}_1$ and ${\cal M}_2$ are scattering amplitudes for processes given in FIG. 4 (a) and (b) respectively. These contributions are
\bea
{\cal M}_1&=&{\cal M}_{01}+{\cal M}_{\beta 1}\label{M11},\\
{\cal M}_2&=&{\cal M}_{02}+{\cal M}_{\beta 2}\label{M21},
\eea
where ${\cal M}_{0i}$ and ${\cal M}_{\beta i}$ are zero and finite temperature parts of the amplitude respectively. Thus
\bea
{\cal M}_1=\bar{u}_3\left[-\frac{i\hbar c \kappa}{4}(\gamma^\mu p_1^\nu+p_3^\mu \gamma^\nu)\right]{u}_1\,D_{\mu\nu\alpha\zeta}(q)\, \bar{u}_4\left[-\frac{i\hbar c \kappa}{4}(\gamma^\alpha p_2^\zeta+p_4^\alpha \gamma^\zeta)\right]{u}_2,
\eea
where $D_{\mu\nu\alpha\zeta}(q)$ is the graviton propagator given in eq. (\ref{graviton}). Zero temperature transition amplitude is
\bea
{\cal M}_{01}=-\frac{\hbar^2 c^2\kappa^2}{32(p_1-p_3)^2}\left[\bar{u}_3(\gamma^\mu p_1^\nu +p_3^\mu\gamma^\nu)u_1\right]\tau\{\bar{u}_4[\gamma_\mu(p_2+p_4)_\nu+\gamma_\nu(p_2+p_4)_\mu -\eta_{\mu\nu}(\slashed{p}_2+\slashed{p}_4)]u_2\},
\eea
and finite temperature transition amplitude is
\bea
{\cal M}_{\beta 1}&=&\frac{\hbar^2 c^2\kappa^2}{16}\left[\bar{u}_3(\gamma^\mu p_1^\nu +p_3^\mu\gamma^\nu)u_1\right]\frac{i\pi\delta(q^2)}{e^{\beta q_0}-1}\left( \begin{array}{cc}1&e^{\beta q_0/2}\\e^{\beta q_0/2}&1\end{array} \right)\times\nonumber\\
&\times&\{\bar{u}_4[\gamma_\mu(p_2+p_4)_\nu+\gamma_\nu(p_2+p_4)_\mu -\eta_{\mu\nu}(\slashed{p}_2+\slashed{p}_4)]u_2\}.
\eea
Then eq. (\ref{M11}) is written as
\bea
{\cal M}_1&=&-\frac{\hbar^2 c^2\kappa^2}{16}\left[\bar{u}_3(\gamma^\mu p_1^\nu +p_3^\mu\gamma^\nu)u_1\right]\{\bar{u}_4[\gamma_\mu(p_2+p_4)_\nu+\gamma_\nu(p_2+p_4)_\mu -\eta_{\mu\nu}(\slashed{p}_2+\slashed{p}_4)]u_2\}\times\nonumber\\
&\times&\left[\frac{\tau}{2(p_1-p_3)^2}-\frac{i\pi\delta(q^2)}{e^{\beta q_0}-1}\left( \begin{array}{cc}1&e^{\beta q_0/2}\\e^{\beta q_0/2}&1\end{array} \right)\right].
\eea
And eq. (\ref{M21}) is 
\bea
{\cal M}_2&=&-\frac{\hbar^2 c^2\kappa^2}{16}\left[\bar{u}_4(\gamma^\mu p_1^\nu +p_4^\mu\gamma^\nu)u_1\right]\{\bar{u}_3[\gamma_\mu(p_2+p_3)_\nu+\gamma_\nu(p_2+p_3)_\mu -\eta_{\mu\nu}(\slashed{p}_2+\slashed{p}_3)]u_2\}\times\nonumber\\
&\times&\left[\frac{\tau}{2(p_1-p_4)^2}-\frac{i\pi\delta(q^2)}{e^{\beta q_0}-1}\left( \begin{array}{cc}1&e^{\beta q_0/2}\\e^{\beta q_0/2}&1\end{array} \right)\right].
\eea

\subsection{Gravitational Compton scattering}

The gravitational Compton scattering is analyzed and following diagrams are considered.
\begin{figure}[ht]
\includegraphics[scale=0.35]{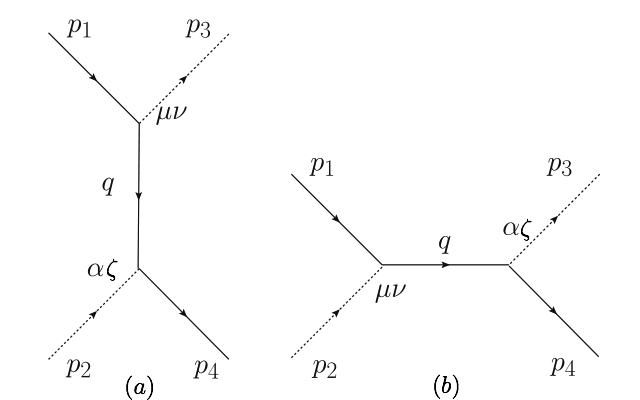}
\caption{Feynman diagram for Compton scattering}
\end{figure}

\newpage

The transition amplitude of the gravitational Compton scattering is
 \bea
{\cal M}= {\cal M}_1+{\cal M}_2,
\eea
where ${\cal M}_1$ and ${\cal M}_2$ are scattering amplitudes for processes given in FIG. 5 (a) and (b) respectively.
\bea
 {\cal M}_1&=&\bar{u}_4\xi^*_{3\mu\nu}\left[-\frac{i\hbar c\kappa}{4}(\gamma^\mu p_1^\nu +(p_1-p_3)^\mu \gamma^\nu)\right]\,S(q)\,\left[-\frac{i\hbar c\kappa}{4}(\gamma^\alpha (p_4-p_2)^\zeta +p_4^\alpha \gamma^\zeta)\right]u_1\xi_{2\alpha\zeta},\\
 {\cal M}_2&=&\bar{u}_4\xi^*_{3\mu\nu}\left[-\frac{i\hbar c\kappa}{4}(\gamma^\mu p_1^\nu +(p_1+p_2)^\mu \gamma^\nu)\right]\,S(q)\,\left[-\frac{i\hbar c\kappa}{4}(\gamma^\alpha (p_3+p_4)^\zeta +p_4^\alpha \gamma^\zeta)\right]u_1\xi_{2\alpha\zeta},
\eea
where $S(q)$ is the fermion propagator given in eq. (\ref{fermion}). Thus transition amplitude are
\bea
 {\cal M}_1&=&-\frac{\hbar^2 c^2\kappa^2}{16}\left[\bar{u}_4(2\vec{p}_4-\vec{p}_2)\cdot \vec{\xi}_2\slashed{\xi}_2\right]\left[\slashed{\xi}_3^* \vec{\xi}_3^*\cdot(2\vec{p}_1-\vec{p}_3)u_1\right]\left[\frac{(\slashed{p}_1-\slashed{p}_3+m)}{(p_1-p_3)^2-m^2}-{\cal F}(\beta)\right],
\eea
and
\bea
{\cal M}_2&=&-\frac{\hbar^2 c^2\kappa^2}{16}\left[\bar{u}_4(2\vec{p}_4+\vec{p}_3)\cdot \vec{\xi}_3^*\slashed{\xi}_3^*\right]\left[\slashed{\xi}_2 \vec{\xi}_2\cdot(2\vec{p}_1+\vec{p}_2)u_1\right]\left[\frac{(\slashed{p}_1+\slashed{p}_2+m)}{(p_1+p_2)^2-m^2}-{\cal F}(\beta)\right],
\eea
where
\bea
{\cal F}(\beta)&\equiv& \frac{2\pi i}{e^{\beta q_0}+1}\Biggl[\frac{\gamma^0\epsilon-\vec{\gamma}\cdot\vec{q}+m}{2\epsilon}\left( \begin{array}{cc}1&e^{\beta q_0/2}\\e^{\beta q_0/2}&-1\end{array} \right)\delta(q_0-\epsilon)\nonumber\\
&+&\frac{\gamma^0\epsilon+\vec{\gamma}\cdot\vec{q}+m}{2\epsilon}\left( \begin{array}{cc}-1&e^{\beta q_0/2}\\e^{\beta q_0/2}&1\end{array} \right)\delta(q_0+\epsilon)\Biggl].\label{F(beta)}
\eea

The graviton polarization tensor $\epsilon_{\mu\nu}$ is taken as the product of two spin-1 polarization vectors $\epsilon_\mu$ and $\epsilon_\nu$, i.e., $\epsilon_{\mu\nu}=\epsilon_\mu\epsilon_\nu$.

\subsection{Graviton photoproduction amplitudes}

The photoproduction of gravitons, such as Born and four-point interaction diagrams, are considered.

\subsubsection{Born diagram}

The Feynman diagram that describes this process is given as follows. 
\begin{figure}[ht]
\includegraphics[scale=0.3]{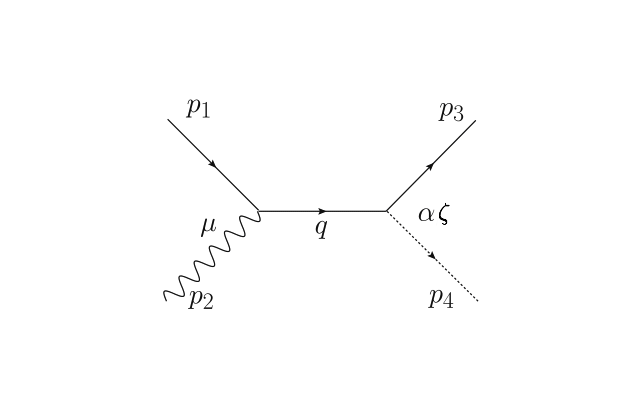}
\caption{Graviton photoproduction: Born diagram}
\end{figure}

The transition amplitude is written as
\bea
{\cal M}=\bar{u}_3\,\xi^*_{4\alpha\zeta}\left[ie\gamma^\mu\right]\xi_{2\mu}\,S(q)\,\left[-\frac{i\hbar c\kappa}{4}(\gamma^\alpha (p_1+p_2)^\zeta +p_3^\alpha\gamma^\zeta)\right]u_1.
\eea
Using the fermion propagator $S(q)$ given in eq. (\ref{fermion}) we get
\bea
{\cal M}=\frac{e\hbar c\kappa}{4}\left[\bar{u}_3(\vec{p}_1+\vec{p}_2+\vec{p}_3)\cdot \vec{\xi}^*_4\slashed{\xi}_4\right]\left[\slashed{\xi}_2 u_1\right]\left[\frac{(\slashed{p}_1+\slashed{p}_2+m)}{(p_1+p_2)^2-m^2}+{\cal F}(\beta)\right],
\eea
where ${\cal F}(\beta)$ is given in eq. (\ref{F(beta)}).

\subsubsection{Four-point interaction diagram}

The second process which describes graviton photoproduction is represented by 
\begin{figure}[h]
\includegraphics[scale=0.3]{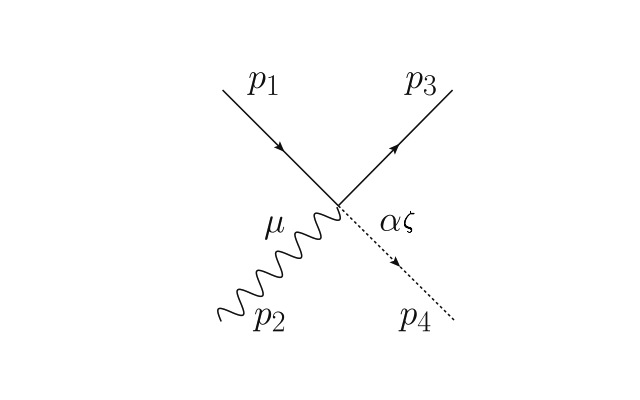}
\caption{Graviton photoproduction: four-point interaction}
\end{figure}

The transition amplitude for this scattering is
\bea
{\cal M}&=&\bar{u}_3\,\xi^*_{4\alpha\zeta}\left[\frac{ie\kappa}{2}\gamma^\zeta\eta^{\alpha\mu}\right]\xi_{2\mu}u_1\nonumber\\
&=&\frac{ie\kappa}{2}\left[\bar{u}_3\slashed{\xi}^*_4\vec{\xi}^*_4\cdot\vec{\xi}_2 u_1\right].
\eea
In this process, there is no temperature contribution, i.e., the temperature does not affect this interaction.

\section{Conclusions}

The theory of gravitoelectromagnetism arises from comparisons between the Newtonian gravity and the Coulomb law. A possible theory for GEM, a gravity theory similar to electromagnetic theory, emerges when the Weyl tensor is considered. The Weyl tensor is divided into two parts, gravitoelectric and gravitomagnetic fields, and the field equations are similar to Maxwell equations. A gravitoelectromagnetic tensor potential leads to a lagrangian formalism. The graviton field is described analogous to electromagnetism and thus provides us with an alternative way to study the interaction of the graviton with fermions and photons in flat space-time. It leads to a perturbation series and transition amplitudes for various scattering processes are considered. GEM at finite temperature using TFD formalism is established.

The graviton, fermion and photon propagators in the TFD formalism are written in two parts, one at T=0 and the other at finite temperature. Transition amplitudes for the gravitational M$\varnothing$ller and Compton scattering and graviton photoproduction at finite temperature are calculated. The transition amplitudes are consist of two parts. These results will have implications for astrophysical processes.

\section*{Acknowledgments}

This work by A. F. S. is supported by CNPq projects 476166/2013-6 and 201273/2015-2. We thank Physics Department, University of Victoria for access to facilities.

\end{document}